\documentclass[conference]{IEEEtran}
\IEEEoverridecommandlockouts
\usepackage{cite}
\usepackage{amsmath,amssymb,amsfonts}
\usepackage{algorithmic}
\usepackage{graphicx}
\usepackage{textcomp}
\usepackage{xcolor}
\usepackage{xspace}
\usepackage{listings}
\usepackage[inline]{enumitem}
\usepackage{xspace}
\usepackage{hyperref}
\usepackage{microtype}
\usepackage{dirtytalk}
\usepackage{makecell}
\usepackage{booktabs}
\usepackage{tabularx}
\usepackage{array}
\usepackage{multirow}
\usepackage{tcolorbox}
\usepackage{colortbl}
\usepackage{stackengine}

\newcommand{\etal}{et al.\@\xspace}
\newcommand{\tool}{Breaking-Good\xspace}
\newcommand{\TODO}[1]{\textcolor{red}{#1}\GenericWarning{}{LaTeX Warning: TODO: #1}}\newcommand\todo\TODO

\newcommand{\javaIncompCategory}{\textbf{Java version incompatibility}\xspace}
\newcommand{\wErrorCategory}{\textbf{Werror failure}\xspace}
\newcommand{\directCategory}{\textbf{Direct compilation failure}\xspace}
\newcommand{\indirectCategory}{\textbf{Indirect compilation failure}\xspace}

\newcommand{\nJavaIncomp}{78\xspace}
\newcommand{\nWerror}{8\xspace}
\newcommand{\nDirectCompilation}{116\xspace}
\newcommand{\nIndirectComp}{38\xspace}
\newcommand{\percentageJavaIncomp}{34}
\newcommand{\percentageWerror}{3}
\newcommand{\percentageDirectCompilation}{47}
\newcommand{\percentageIndirectCompilation}{16}

\newcommand{\totalPercentage}{70}

\def\BibTeX{{\rm B\kern-.05em{\sc i\kern-.025em b}\kern-.08em
    T\kern-.1667em\lower.7ex\hbox{E}\kern-.125emX}}

\newtheorem{definition}{Definition}

\begin{document}

\setlength{\abovecaptionskip}{1mm plus 0mm minus 0mm}
\makeatletter
\newcommand{\linebreakand}{%
  \end{@IEEEauthorhalign}
  \hfill\mbox{}\par
  \mbox{}\hfill\begin{@IEEEauthorhalign}
}
\makeatother

\title{\tool: Explaining Breaking Dependency Updates with Build Analysis}

\author{\IEEEauthorblockN{1\textsuperscript{st} Frank Reyes}
\IEEEauthorblockA{\textit{KTH Royal Institute of Technology}\\
Stockholm, Sweden \\
frankrg@kth.se}
\and
\IEEEauthorblockN{2\textsuperscript{th} Benoit Baudry}
\IEEEauthorblockA{\textit{Université de Montréal}\\
Montréal, Canada \\
baudry@kth.se}
\and
\IEEEauthorblockN{3\textsuperscript{th} Martin Monperrus}
\IEEEauthorblockA{\textit{KTH Royal Institute of Technology}\\
Stockholm, Sweden \\
monperrus@kth.se}

}
\maketitle

\begin{abstract}
Dependency updates often cause compilation errors when new dependency versions introduce changes that are incompatible with existing client code. 
Fixing breaking dependency updates is notoriously hard, as their root cause can be hidden deep in the dependency tree.
We present \tool, a tool that automatically generates explanations for breaking updates.
\tool provides a detailed categorization of compilation errors, identifying several factors related to changes in direct and indirect dependencies, incompatibilities between Java versions, and client-specific configuration.
With a blended analysis of log and dependency trees, \tool generates detailed explanations for each breaking update. These explanations help developers understand the causes of the breaking update, and suggest possible actions to fix the breakage.
We evaluate \tool on 243 real-world breaking dependency updates.
Our results indicate that \tool accurately identifies root causes and generates automatic explanations for \totalPercentage\% of these breaking updates.
Our user study demonstrates that the generated explanations help developers. 
\tool is the first technique that automatically identifies the causes of a breaking dependency update and explains the breakage accordingly. 
\end{abstract}


\section{Introduction}

Dependency management and maintenance is an essential task in modern software development \cite{Prana2021OutProjects}.
Projects can have tens or hundreds of dependencies, which makes the maintenance task hard and time consuming \cite{Pashchenko2020AImplications}.
One challenge in maintaining dependencies up-to-date is that the new dependency version can introduce incompatible changes that prevent the compilation of the client application \cite{VassalloEveryResolution}.
For example, a new version of the dependency may change the API of the used library and break the build of the client code \cite{ruiz2015beyond,dagenais2011recommending}.
Fixing these dependency related errors is a complex task that demands a detailed joint analysis of the code and its dependencies, along with a clear understanding of the expected behavior \cite{10.1145/3236024.3275535}.

Several studies have analyzed the errors causing breaking dependency updates \cite{Jayasuriya2024UnderstandingApplications,Ochoa2022BreakingStudy}.
It has been shown that the large amount of information generated in logs during client compilation, along with the need to differentiate between warnings and critical errors, poses significant challenges for developers \cite{10.1145/3338906.3338917}.
Overall, understanding the causes of a breaking update only from logs consumes valuable time and reduces the efficiency of software maintenance \cite{Goncalves2022DoLoad}.

To address this problem, \emph{we propose to generate explanations for breaking updates}. The explanations should state the underlying causes of the breaking update point to the elements involved in the breakage.
The generating explanations would allow developers to reduce the time spent analyzing large volumes of log data and to quickly identify and fix the root cause of the breaking dependency update.

In this study, we present \tool, a tool to automatically generate explanations for breaking updates.
\tool is founded on a comprehensive categorization of compilation errors related to changes in direct dependencies and indirect dependencies, incompatibilities between Java versions, and client-specific configurations.
\tool extracts information from logs, analyzes dependency trees, and parses the application code to craft the best possible explanation.
Using this extracted information, \tool first categorizes the causes of the breakage.
For each category of compilation error, \tool provides an explanation template, ensuring that each explanation highlights the cause of the compilation error and provides details relevant to its resolution.
The \tool explanations steer the developers' attention toward the causes of the breaking dependency updates and provide suggestions for possible actions to fix the problem.

We evaluate \tool using a dataset of 243 breaking dependency updates in real-world projects.
Our results indicate that \tool successfully identifies the root causes of compilation errors in \totalPercentage\% of cases.
In particular, \tool determines that \nIndirectComp breaking dependency updates are due to changes in the indirect dependencies of the project.
A user study enabled us to validate the core structure of the explanation provided to developers. 
To the best of our knowledge, \tool is the first tool that helps developers in mitigating the common problem of breaking dependency updates with generated explanations.

To summarise, we make the following contributions:
\begin{itemize}
    \item A novel approach for explaining breaking updates, which combines the analysis of build logs with the semantic comparison of dependency trees. We provide a complete overview of the dependency-related factors that trigger broken compilation in client code, with rare insights on indirect dependencies causing breaking dependency updates.
    \item \tool, an original tool that automatically generates explanations for breaking updates, in order to help developers find, understand and fix build breakages.
    \tool source code is available in \href{https://github.com/chains-project/breaking-good}{https://github.com/chains-project/breaking-good}
    \item A quantitative evaluation and a user study to evaluate \tool. We run the tool on 243 real dependency updates that caused compilation errors in the field, proving its effectiveness in \totalPercentage\% of the cases. The interviewed developers validate the usefulness of the explanations generated by \tool.
\end{itemize}

\section{Method for Explaining Breaking Dependency Updates}
In this section, we introduce the key concepts used in this work and we outline the main steps to automatically generate explanations for breaking dependency updates.

\subsection{Concepts}

The concept of \textit{breaking dependency update} is a well known pain for practitioners. For a client project that declares a set of third-party dependencies, a breaking dependency update refers to an update of a dependency version, which results in breakage of the project's build \cite{reyes2024bump}. 

\begin{definition}
    \label{breakdepupdate}
    A \textbf{breaking dependency update} is a pair of commits for a project composed by a pre-breaking commit with a passing build and a breaking commit that updates the version of one single dependency and fails the build.
\end{definition}

In \autoref{fig:bdu_example} we illustrate a breaking dependency update found in the 
\href{https://github.com/knaufk/flink-faker/pull/52/files}{flink-faker} project. We see a commit which updates the \texttt{datafaker} dependency from version 1.3.0 to version 1.4.0.
When the continuous integration tries to build after this commit, the build fails due to compilation error \emph{incompatible types: java.util.Date cannot be converted to java.sql.Timestamp}.

\begin{figure}[htp]
    \centering
    \href{https://xkcd.com/2347/}{\includegraphics[width=0.80\linewidth]{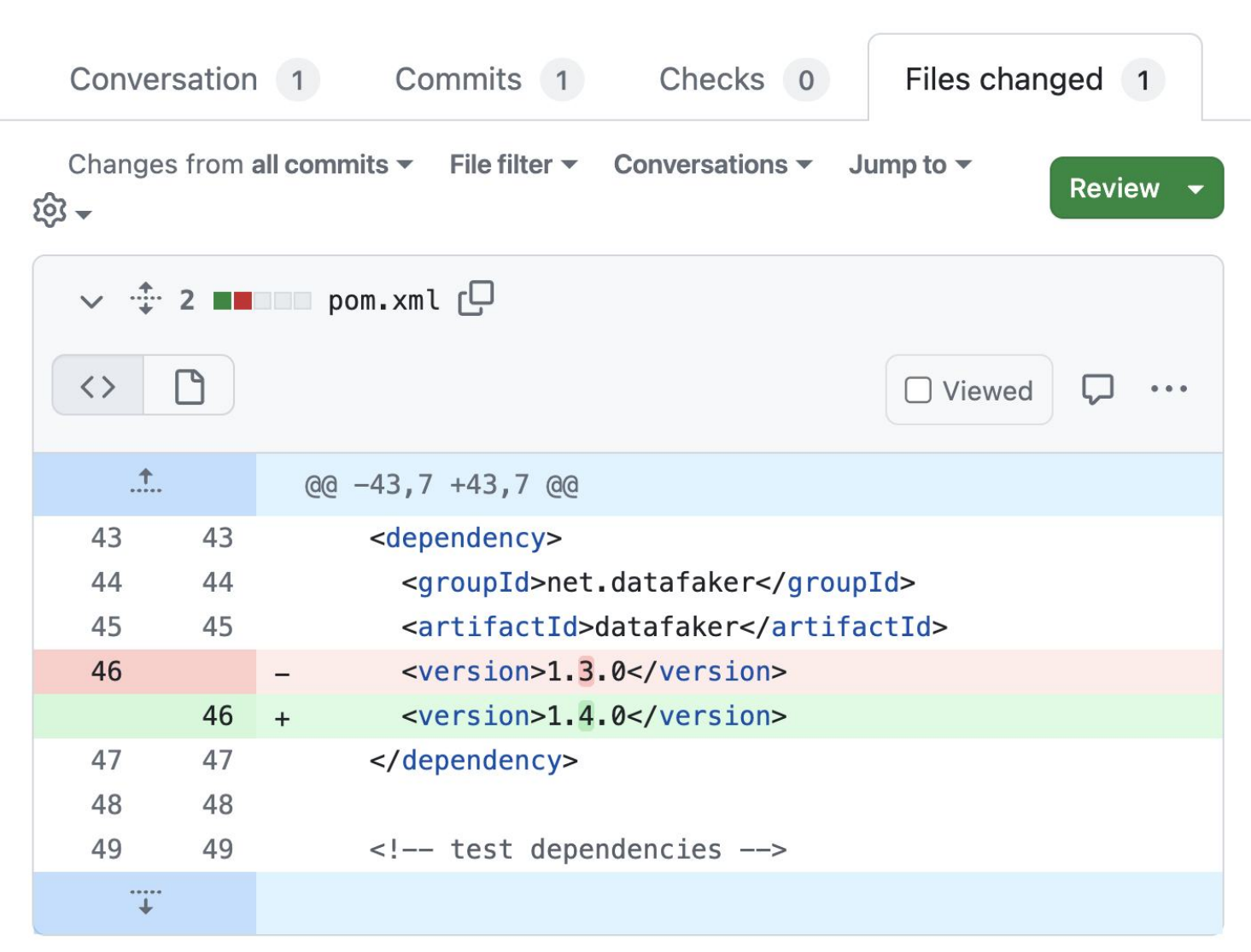}}
    \caption{Commit with dependency version update in Maven.}
    \label{fig:bdu_example}
\end{figure}

In this work, we aim at generating a human readable and actionable explanation that clarifies why and how the dependency update breaks the build and what the possible actions to fix the breakage are.

To effectively generate explanations, we need to know which code elements have changed between the two versions of the dependency, and how these elements interact with the client project's code.
Java to basic elements such as classes, methods, and variables as \texttt{construct} \cite{TheSpecification}.
Here, we use this term to designate any element that can change in the new dependency version. 

\begin{definition}
    \label{construct}
    A \textbf{construct} is a basic building unit in software, such as a class, function, module, and other similar elements. A dependency update can add or remove \textbf{constructs}, or modify their interface or behavior.
\end{definition}

\begin{figure*}[ht]
    \centering
    \includegraphics[width=1\linewidth]{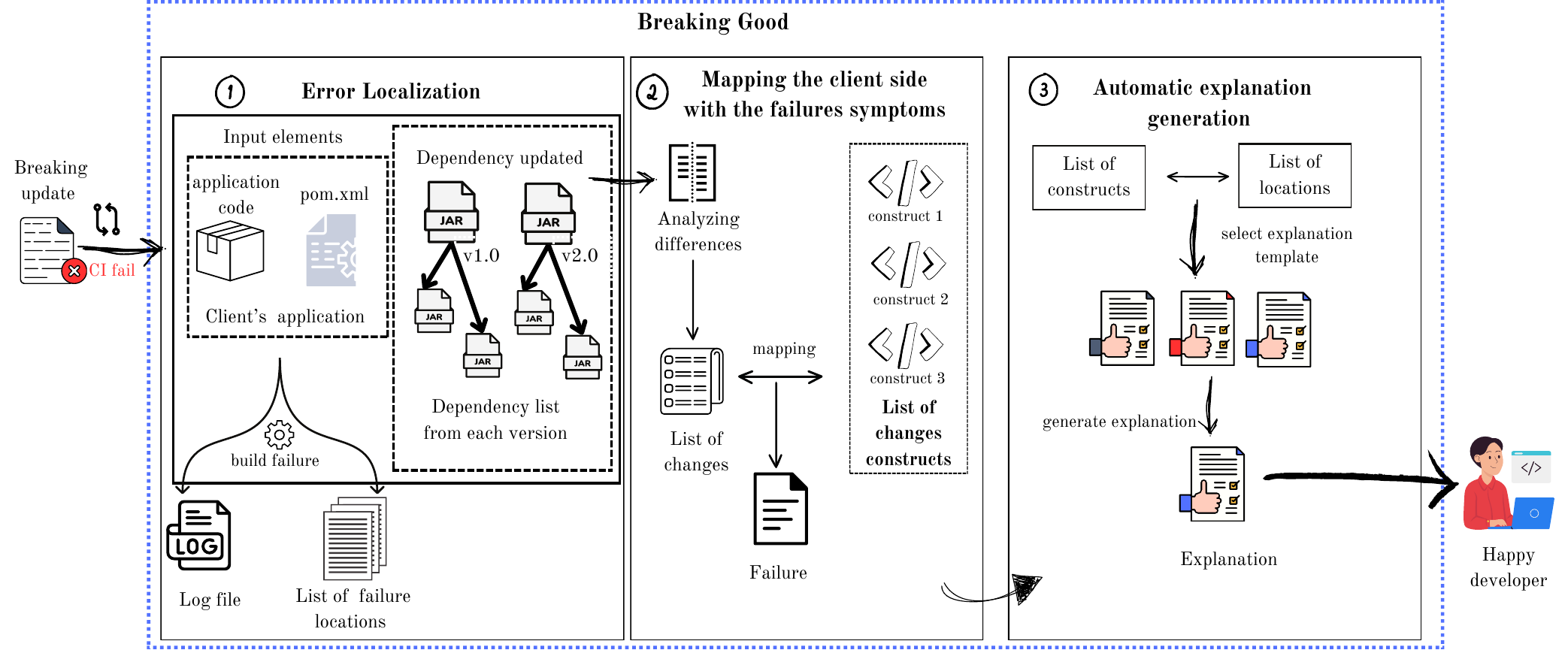}
    \caption{Overview of \tool to automatically generate explanations for breaking dependency updates.}
    \label{fig: methodology}
\end{figure*}

\subsection{Overview}

Our process to automatically generate explanations for breaking dependency updates starts with three inputs.
It takes both versions of the updated dependency: the original version, which acts as the stable reference, and the updated version.
The analysis and comparison of both versions will provide the list of constructs that have changed in the new version.
The third input is the source code of the client project.
The analysis of the client code, with respect to the changed constructs, will provide indications about the possible locations that need to be fixed to address the build breakage.

As output, we generate an explanation for the breakage.
The explanation includes information about the causes of the build failure.
We also generate suggestions on how to fix the failure in the client's project code. For the \href{https://github.com/knaufk/flink-faker/pull/52/files}{flink-faker} breaking update example in \autoref{fig:bdu_example}, we would automatically generate the explanation shown in \autoref{fig:explanation}

\begin{figure}[hpt]
    \centering    
    \href{https://github.com/chains-project/breaking-good/blob/main/Explanations/1ef97ea6c5b6e34151fe6167001b69e003449f95.md}{\includegraphics[width=.80\linewidth]{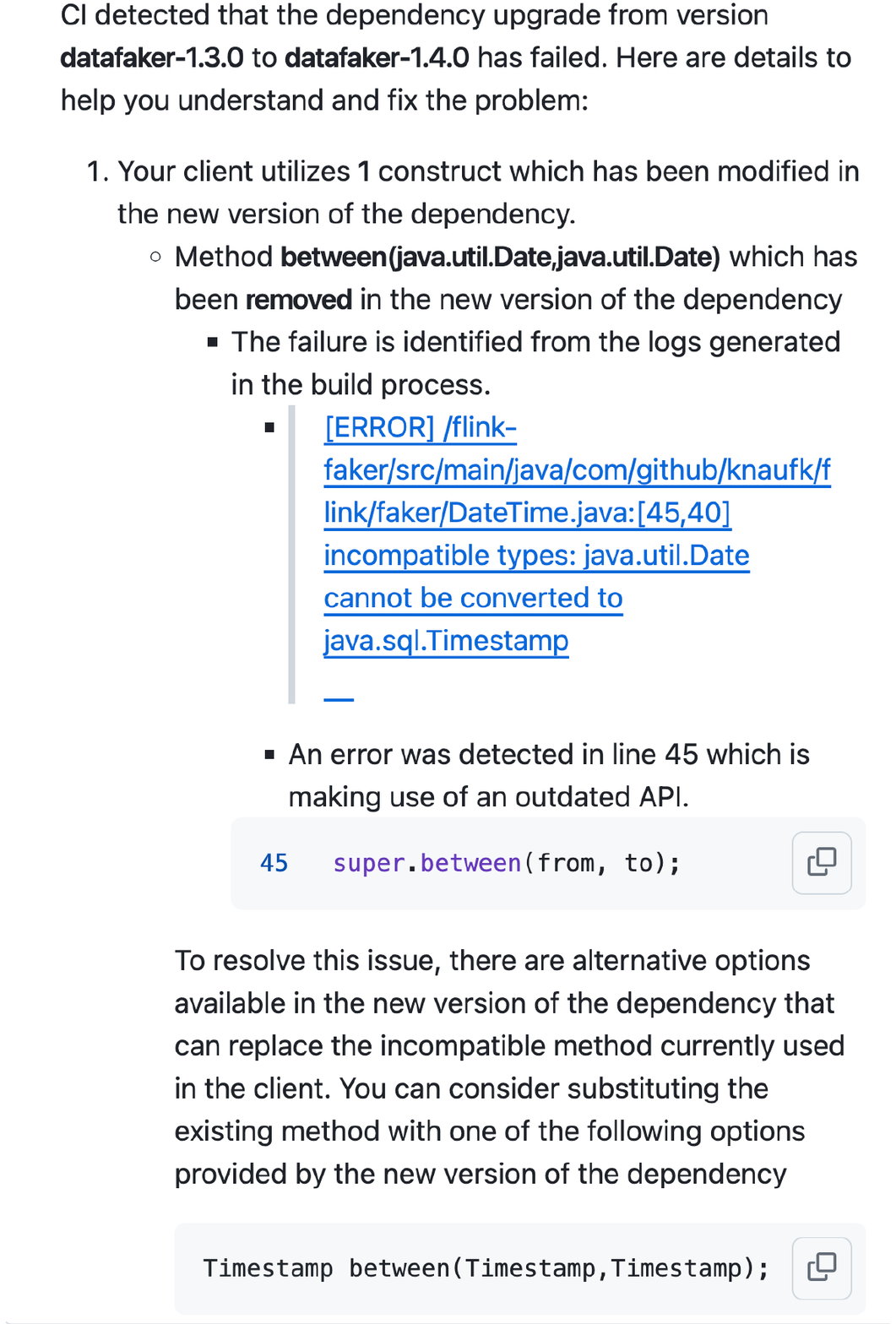}}
    \caption{Generated explanation for the breaking dependency update in \texttt{flink-flaker} project.}

    \label{fig:explanation}
\end{figure}

In \autoref{fig: methodology}, we show the different steps to automate the generation of explanations.
The process starts with two independent actions.
On the one hand, we analyze the logs generated by the build system to determine the type of error caused by the dependency update (step 1).
On the other hand, we enumerate the list of constructs that have been modified between the two versions of the dependency (step 2).
Then, we combine the results of both analyses to identify the client elements affected by the changes in the new dependency version (step 3), and we automatically generate the explanation for the breaking dependency update (step 4).
Each step is discussed in the following sections.

\subsection{Error localization in client code}
\label{sec:error-loca}

We locate the source of each compilation error in the log files generated by the client's build.
Error lines are matched in the log to extract various types of information, including warnings, file paths, error codes, and even the actual line in the client code.

For instance, consider the following error line: \texttt{[ERROR] /flink-faker/src/main/java/com/github/
knaufk/flink/faker/DateTime.java:[45,40] incompatible types: java.util.Date cannot be converted to java.sql.Timestamp}.
From this line, we extract information such as the file path 
\texttt{(/flink-faker/src/main/java/com/github/
knaufk/flink/faker/DateTime.java)} and the specific location in the file \texttt{([45,40])}. This process allows us to precisely identify the nature and context of each error.

This classification step outputs, a list of errors caused by breaking dependency updates, as well as detailed information about the failure.

\subsection{Categories of breaking updates}
\label{BG-categories}

There is a wide variety of build failures \cite{Hassan2019TacklingIntegration}.
To generate an appropriate explanation, we first need to categorize the type of build failure. This will inform the type of explanation we generate. 
First, we focus on compilation errors and discard test failures due to change behavior.

After a systematic manual analysis of real-world dependency updates, we identify four categories of compilation errors: \textbf{Werror failure}, \textbf{Java version incompatibility}, \textbf{Direct compilation failure}, and \textbf{Indirect compilation failure}.

A build failure is categorized as a \wErrorCategory when the compiler treats warnings as errors. In Maven, this option is \texttt{-Werror}, which stands for `warning as error'.
When the new version of the dependency introduces changes that generate warnings in the client project's code, this results in a build breakage.

A build failure is considered a \javaIncompCategory if the Java bytecode version used for a dependency is incompatible with the JVM used during the compilation of the client project.
This may cause bytecode incompatibilities or missing APIs that have been removed or introduced in some JVM versions.

The breaking update is categorized as a \directCategory if the log file contains the pattern \say{COMPILATION ERROR} and the project's compilation fails.
This failure occurs when the updated dependency directly causes compilation errors in the client's code.

In addition, a breaking update is considered a \indirectCategory if the cause of the failure is due to changes in the code of an indirect dependency (and not of the bumped dependency).
We identify these failures when no changes can be found in the updated dependency.
Yet, we find the root cause by analyzing the differences between the lists of indirect dependencies of both versions of the updated dependency.

\autoref{tab:categories} summarizes the four main categories.

\begin{table}[hbt]
\caption{List of breaking update categories.}
\centering
\footnotesize
\label{tab:categories}
\begin{tabularx}{\columnwidth}{X X}
\toprule
Compilation failure category & \makecell{Definition} \\
\midrule
  \multirow{3}{*}{Werror failure} & The compiler treats warnings as errors due to the -Werror option enabled in the build configuration file.\\
  \hline
  \multirow{4}{*}{Java version incompatibility} & The Java version required by a dependency is incompatible with the JVM version used during the client project's build. \\
  \hline
  \multirow{3}{*}{Direct compilation error} & The updated dependency directly causes compilation errors in the client's code. \\
  \hline
  \multirow{4}{*}{Indirect compilation error} & The client code uses APIs from an indirect dependency that have been updated, added, or removed, causing ripple compilation errors in the client under consideration. \\
 \bottomrule
\end{tabularx}
\end{table}

Categorizing the breakage is essential for explaining the root causes of the breakage to the developer.

\subsection{Mapping the client side errors with the root causes in dependencies}

To identify the causes of the breaking update, we analyze the relation between the compilation errors in the client project and various factors that could trigger these failures.
The causes can include changes in the new version of the dependency, incompatibilities in Java versions, conflicts with indirect dependencies, or specific errors due to particular configurations.

First, we analyze the compilation logs that contain the dependency update errors.
We use regular expressions to identify patterns which allow us to detect the category of failures (\autoref{BG-categories}).

To determine if the cause of the failure is due to \javaIncompCategory, we search for the predefined keyword \say{class file has wrong version} in the logs.
For example, when the dependency \texttt{spring-context} is updated from version \texttt{5.3.23} to version \texttt{6.0.2} in the \texttt{camunda-platform-7-mockito} project, the build fails due to the new version of the dependency being incompatible with the Java 11 version in which the project is built.

We determine the cause \wErrorCategory failure if the logs contain the keywords \say{class file has wrong version}. 
For instance, by updating the \texttt{hibernate-entitymanager} dependency from version \texttt{4.2.11.Final} to \texttt{5.5.14.Final} in the \texttt{nem} project the compilation process fails because the \texttt{-Werror} option is enabled in the configuration file. 
The new version of the dependency flags as deprecated the \texttt{Query} interface which is invoked in the client code generating warnings and causing the compilation error.

To identify if the failure is due to \directCategory, first, we identify the constructs present in the client lines of code that cause compilation errors (after error localization, see \autoref{sec:error-loca}), resulting in a list of constructs.
Next, we analyze whether some of the identified constructs are present in the list of constructs that have changed in the new version of the dependency.
To get this list of changed constructs, we use state of the art API differencing \cite{Japicmp-baseJapicmp}, the result is a list of constructs that have changed in the new dependency version.
For example, when updating the \texttt{datafaker} dependency from version \texttt{1.3.0} to version \texttt{1.4.0} the client build fails due to compilation errors.
We identify from the lines that trigger the error the following constructs: \texttt{Timestamp}, \texttt{between}, \texttt{getTime}.
As a result, we determine that the compilation error is because the construct \texttt{between} has been modified in the new version of the dependency, causing the compilation error.

To identify \indirectCategory as the cause of the failure we retrieve the complete list of indirect dependencies for the new and previous versions of the updated dependency and compare both lists.
From this comparison, each indirect dependency is in one of the following situations: no changes, deleted, added, or updated.

If a dependency has been added or removed from the list of indirect dependencies, then we analyze whether the breakage is due to the use of deleted dependency constructs or incompatible constructs defined in the new indirect dependencies. 
For instance, the project \texttt{docker-adapter} build fails to update the \texttt{http} dependency from version \texttt{v0.25} to version \texttt{v1.1.0} because the \texttt{org.cactoos} dependency was removed from version \texttt{v1.1.0}.

When some indirect dependencies have been updated in the new version, we analyze which constructs change in the new version of each updated indirect dependency.
As a result, we obtain a complete list of all modified constructs.
Subsequently, we examine whether the components causing the breakage update belong to any of the indirect dependencies that have been updated.
For example, the update of \texttt{logback-classic} from version \texttt{1.2.11} to version \texttt{1.4.1} in the \texttt{pay-adminusers} project fails the build because it is not possible to load the \texttt{org.slf4j.spi.LoggingEventAware} class from the indirect dependency \texttt{slf4j-spi}

To recapitulate, we propose a novel process of mapping the information between the error information from the build logs and the differencing of updated dependencies.
As a result, we have a sound and robust categorization of the breakage cause per the ones of \autoref{BG-categories}.

\subsection{Generate an explanation with templates}
\begin{figure}[hpt]
    \centering
    \href{https://github.com/chains-project/breaking-good/blob/main/Explanations/0cdcc1f1319311f383676a89808c9b8eb190145c.md}{\includegraphics[width=.90\linewidth]{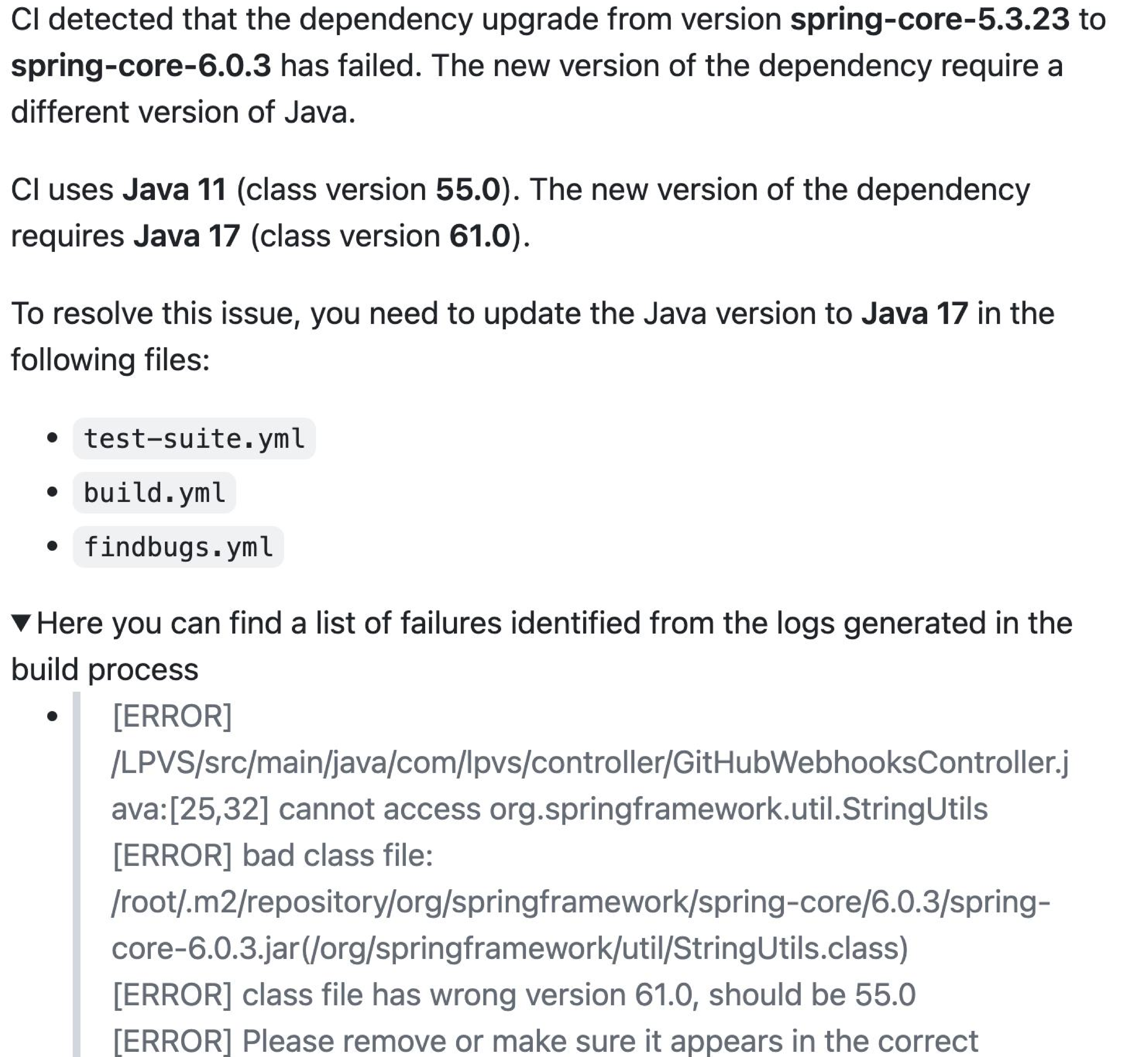}}
    \caption{Generated explanation for a breaking dependency update due to \javaIncompCategory.}
    \label{fig:java_version_incompatibility_exp}
\end{figure}

To generate explanations, we have defined specific templates for each breakage category.

\textbf{Java incompatibility failures} are explained by specifying which Java versions are required by the client to compile with the new dependency version.
The explanation indicates which workflow files need to be updated, with the actual Java version required by the new version of the dependency to solve the version incompatibility problem.
In addition, the explanation lists the errors extracted from the logs that are triggered by the incompatibility problem.
\autoref{fig:java_version_incompatibility_exp}
show an example of the explanation.

For \textbf{Werror Failures}, the explanation template flags warnings that have been escalated to errors due to the \emph{Werror} option.
As a result, it presents the list of warnings generated by the use of deprecated constructs which belong to the new version of the dependency updated used in the client application.
An example explanation can be found in the \autoref{fig:werror_explanation_example} 
\begin{figure}[hpt]
    \centering    
    \href{https://github.com/chains-project/breaking-good/blob/main/Explanations/c5fd5187ce64d2b53602717f09cc18dd21d55e8d.md}{\includegraphics[width=.80\linewidth]{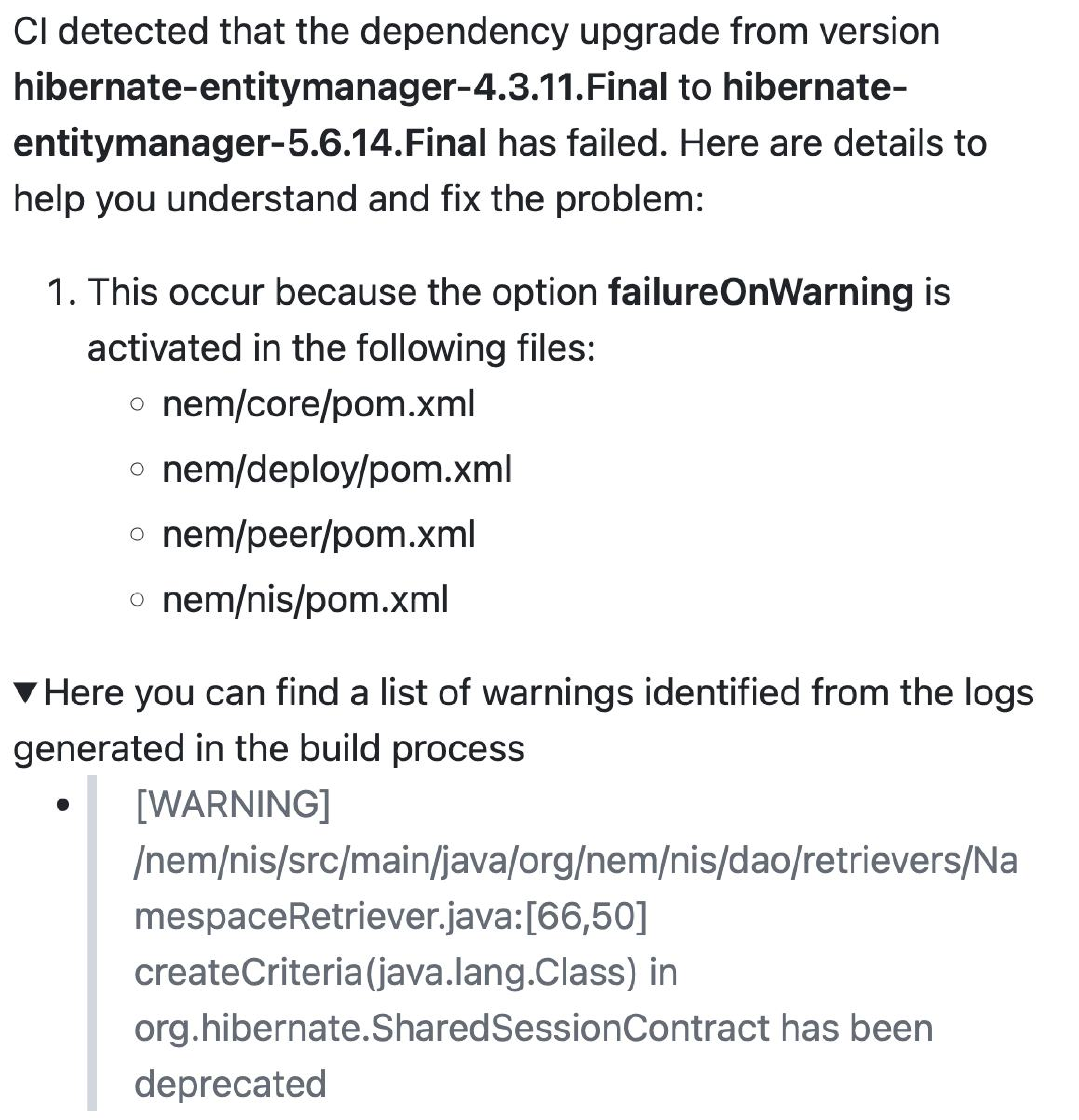}}
    \caption{Generated explanation for a breaking dependency update due to \wErrorCategory.}
    \label{fig:werror_explanation_example}
\end{figure}

In the case of \textbf{Direct compilation failures}, the template highlights the number of constructs causing the compilation failures.
For each construct, the template specifies its signature, the lines where it is used in the client code, and the information gathered from the logs related to the failure.
In addition, it includes suggestions for constructs added in the new version of the dependency that may be utilized to fix the problem.
 To create this list of suggestions, first, we identify the constructs added in the updated version of the dependency.
 Next, we analyze such constructs that have the same name and belong to the same class as the construct that causes the breakage
For example, \autoref{fig:explanation} is an explanation generated from a case of direct compilation failures. 

For \textbf{Indirect compilation failure}, the template follows the same structure as for \textbf{Direct compilation failure} with the number of constructs causing the compilation error and the information gathered from the logs related to the failure. Also, it also includes suggestions for alternative constructs.
Crucially, it also includes additional information about the indirect dependency: 1) its changed version and 2) where it fits in the dependency tree.
An example explanation can be found in the \autoref{fig:indirect_explanation}
\begin{figure}[hpt]
    \centering
    \href{https://github.com/chains-project/breaking-good/blob/main/Explanations/b8f92ff37d1aed054d8320283fd6d6a492703a55.md}{\includegraphics[width=.80\linewidth]{Figure/indirect_explanation.pdf}}
    \caption{Generated explanation for a breaking dependency update due to \indirectCategory.}
    \label{fig:indirect_explanation}
\end{figure}

Each template ensures that each breakage category is addressed with a clear, human-readable explanation, providing developers with the necessary, concrete information to identify and fix the broken build.

\subsection{Integration in a development workflow}

To keep Java project dependencies up to date, developers can choose to update them manually or do automated updates through bots such as Dependabot or Renovate. They automate the process by creating pull requests with the required updates.
Such updates can break the application build due to changes in the API or behavior of the new dependency version.
To address this problem, \tool can be tightly integrated into the dependency update process by monitoring dependency update pull requests, regardless of whether they are generated by bots or humans.
\tool would 1) automatically generate the detailed explanation about the causes of the broken dependency update 2) push that explanation as a comment on the pull request.
These explanations integrated in the development platform allow developers to quickly access the information needed to understand and fix the build breakage.

\subsection{Implementation}

\tool is implemented in Java and runs on Java 17. 
We locate the lines that cause compilation errors in the client code using AST analysis with Spoon\cite{pawlak:hal-01169705}, based on the information gathered from the logs. 
We rely on japicmp\cite{Japicmp-baseJapicmp}, a static analysis tool that detects binary changes between dependencies, to identify changes causing failures.
We use Maven to generate the dependency tree for the new and previous versions of the dependency update.

\section{Experimental Methodology}

In this section, we present the research questions that serve as the basis to evaluate \tool, as well as the methodology to answer them.

\subsection{Research Questions}

In this paper, we study the following research questions:

\newcommand\rqFailureTypes{What is the prevalence of each failure category behind breaking dependency updates?\xspace}
\newcommand\rqExplanation{How helpful are the automatically generated explanations for developers to understand and fix breaking dependency updates?\xspace}
\newcommand\rqTransitiveDep{What is the extent of breaking dependency updates due to indirect dependencies?\xspace}


\begin{enumerate}[label=RQ\arabic*:, ref=RQ\arabic*]

    \item \label{rq: failures}\textbf{\rqFailureTypes}

    \item \label{rq: transitive}\textbf{\rqTransitiveDep}

    \item \label{rq: explanations}\textbf{\rqExplanation}

\end{enumerate}

\subsection{Study Subjects}

We performed all experiments on the BUMP\footnote{\href{https://github.com/chains-project/bump}{https://github.com/chains-project/bump}} benchmark, which contains a total of 571 fully reproducible breaking dependency updates.
These updates are distributed over 153 Java projects collected from GitHub.
Each project has more than 7 contributors, over 100 commits, and more than 10 stars, ensuring significant diversity in the selected projects.
The generation of automatic explanations was applied for the 243 breaking updates categorized as compilation failures by BUMP.

\subsection{Methodology for RQ1}
\label{RQ1Methodology}

The goal of this research question is to determine the occurrence of different types of compilation errors that cause breaking dependency updates, and the prevalence of the explanations generated for each category.
First, we reproduce the breaking dependency updates of BUMP and select those that fail due to compilation errors.
We examine each log generated by the breaking updates to classify the failures.
We categorize the compilation errors into four categories, as presented in \autoref{BG-categories}.
Second, we generated explanations using specific templates for each dependency break update.
We then analyzed the total number of explanations generated and the root causes of the missing explanations.
Through this approach, we provide a detailed and quantitative understanding of the underlying causes of compilation errors resulting from breaking dependency updates.

\subsection{Methodology for RQ2}

A key novelty of our work is to consider that a dependency update can break because of an indirect dependency. In this question, we investigate these cases in detail.
We use the subset of \nIndirectComp breaking updates categorized as \indirectCategory in \textbf{RQ1}.
We analyze whether the indirect dependency was modified, removed, or added in the new version of the updated dependency.
We also analyze how the constructs that trigger the breaking update failure affect the client's compilation process.

\subsection{Methodology for RQ3}

In this research question, we aim to evaluate the quality of explanations generated by \tool.
We designed an online questionnaire to collect feedback about developers.
We selected 5 researchers specialized in software engineering and invited them to participate in the study.
The participants analyzed two different cases of breaking updates.
The questionnaire included both Likert-scale questions and open-ended questions, to obtain a detailed evaluation of the level of difficulty in identifying the causes of breaking updates, the usefulness of the explanations provided, and the strengths and weaknesses of the generated explanation.
The answers to Likert-scale questions were analyzed quantitatively to identify trends and patterns in participants' perceptions.
Responses to open-ended questions were analyzed qualitatively to identify recurring themes and better understand the researchers' feedback.

\section{Experimental Results}

In this section, we present the experimental results to answer our research questions.

\subsection{RQ1 (Breaking Category Prevalence)}

The goal of this RQ is to classify the compilation errors behind breaking dependency updates and the prevalence for each explanation type.
Each compilation error is classified into four different categories, as described in \autoref{tab:categories}. 
In \autoref{tab:compilation-category-counts} we report the number of occurrences of each compilation error category and the number of generated explanations per category. 

\begin{table}[htp]
\caption{Number of breaking updates per breakage category and number of successfully generated explanations per category.}
\centering
\rowcolors{2}{gray!10}{white}
\begin{tabular}{lcc}
\toprule
Compilation error category  &  \makecell{\begin{tabular}{@{}c@{}} Number of \\  breaking updates
  \end{tabular}} & \begin{tabular}{@{}l@{}} Number of \\  explanations
  \end{tabular} \\ 
\midrule
  \directCategory &  \nDirectCompilation (\percentageDirectCompilation\%) & 82 \\
  \indirectCategory &  \nIndirectComp (\percentageIndirectCompilation\%)& 2 \\
  \javaIncompCategory &  \nJavaIncomp (\percentageJavaIncomp\%)  & 78\\
  \wErrorCategory &  \nWerror (\percentageWerror\%)  & 8\\
\bottomrule
 
\end{tabular}
\label{tab:compilation-category-counts}
\end{table}

BUMP contains a total of 243 dependency updates that fail due to compilation errors.
\tool aims to generate automatic explanations for each of the categorized breaking updates in \autoref{tab:compilation-category-counts}.
Each explanation is generated according to the corresponding template.
\tool generates a total of 181 explanations for 243 breaking dependency updates that fail due to compilation errors.

Out of our set of 243 breaking updates, \nDirectCompilation(\percentageDirectCompilation\%) fail due to a change in the updated dependency, which affects the client application. They are categorized as \directCategory.
Such failures occur when the client application uses constructs that have been modified or removed in the new version of the dependency.
The explanations detail the modifications or deletions of constructs in the new version of the dependency and suggest alternatives to resolve the problem.
Continuing with the example in \autoref{fig:bdu_example}, when updating the \texttt{datafaker} dependency from version \texttt{1.3.0} to version \texttt{1.4.0}, the compilation fails due to a change in the API of the \texttt{between} method in the new version of the updated dependency.
\autoref{fig:explanation} contains the generated explanation of the update after the breaking update.
The explanation shows how the change in the API of the \texttt{between} method causes the breaking update.
 
We categorized \nIndirectComp(\percentageIndirectCompilation\%) breaking updates as \indirectCategory.
This happens when an indirect dependency is added, removed, or modified in the dependency tree of the updated dependency and when the indirect dependency is used in the application code.
For example, \autoref{fig:indirect_explanation} shows the dependency update of \texttt{maven-surefire-common} from version \texttt{3.0.0-M5} to version \texttt{3.0.0-M7}.
The upgrade fails due to the update of the indirect dependency \texttt{surefire-api} from version \texttt{3.0.0-M5} to version \texttt{3.0.0-M7}.
The new version of the indirect dependency removes the \texttt{getInstance} method that is used in the client application.

\tool successfully generates an explanation for 82 of the 116 \directCategory, as well as for 13 of the 38 broken updates classified as \indirectCategory.
The explanations provide details about the affected dependencies, and the specific constructs causing the errors. 
They also suggest adjustments to the client code to resolve the problems.
\tool cannot generate an explanation when the comparison between the two versions of a dependency returns no element. 
In these cases, we know that a dependency has changed, we know the dependency is the root cause for the breaking update, but we cannot retrieve the exact set of constructs that have changed between the two versions of the dependency. 
This is due to the underlying tool for comparing the two dependency versions, japicmp, and we leave for future work the improvement of this tool. 
In section \autoref{indirect_dependencies}, we provide a more detailed analysis of the causes that trigger indirect compilation failures.

Java version incompatibility occurs in \nJavaIncomp (\percentageJavaIncomp\%) cases.
This happens when the Java version used in the workflow to compile and test the client application differs from the Java version required by the new version of the dependency.
For this category, \tool automatically generates an explanation for all the \nJavaIncomp breakages.
\autoref{fig:java_version_incompatibility_exp} shows an example of an explanation generated for updating the \texttt{spring-core} dependency from version \texttt{5.3.23} to version \texttt{6.0.3}. The explanation points the developer to the cause of the breakage: an incompatibility between Java 11, which is used to compile the client, and Java 17, which is required by the new version of the dependency.

The least frequent cause of compilation error is a Werror failure.
In this scenario, the build system treats warnings as errors, resulting in compilation failure.
Warnings occur, for example, when the new version of the dependency declares some constructs as deprecated, and they are used in the client code.
We observe this phenomenon for \nWerror (\percentageWerror\%) breaking dependency updates.
\tool generates an explanation for each of these \nWerror breakages.
The generated explanation shows the warnings generated by the deprecated constructs.
It also includes the list of build configuration files that contain the enabled option.
\autoref{fig:werror_explanation_example} shows the explanation generated for the breaking update of the dependency \texttt{hibernate-entitymanager} from version \texttt{4.3.11.Final} to version \texttt{5.6.14.Final} in the project \texttt{nem}.
The explanation highlights how 4 configuration files contain the enabled option causing the breaking updates.

To our knowledge, this experiment is the most fine-grain overview of the most common causes of compilation failures for breaking updates.
Changes in the direct dependency's API only cause half of the failures (\percentageDirectCompilation\%). Java version incompatibility and changes in indirect dependencies are also major sources of breaking dependency updates, a result that has never been discussed before.
This detailed knowledge of root causes allows us to propose high quality explanations to developers in order to locate and fix the issue.
In the case of failures due to Java version incompatibilities and client-specific configurations, \tool is capable of generating 100\% of the explanations for the failure.

\begin{tcolorbox}[boxrule=1pt,arc=.3em, left=4pt, right=4pt]
\textbf{Answer to \ref{rq: failures}}:
On the recent BUMP benchmark \cite{reyes2024bump}, \tool finds that 47\% of compilation errors are due to changes in direct dependencies, 33\% are due to changes in indirect dependencies, 17\% are caused by Java version incompatibility and 4\% result from Werror failures.
\tool automatically generates explanations for \totalPercentage\% of these breaking dependency updates.
These results highlight the diverse nature of compilation errors and clearly show the need for considering different scenarios for explaining build failures due to dependency updates.
\end{tcolorbox}

\subsection{RQ2 (Indirect Dependency Problem)}
\label{indirect_dependencies}

In this research question, we dive into one specific type of breaking dependency update: the ones classified as \indirectCategory. 
This is a special type of breakage, where a project updates one of its dependencies, but the breakage is not due to a change in that direct dependency.
Instead, the breakage is due to an indirect dependency.
Our paper is the first to analyze this problem which happens a lot in the field and poses problems to practitioners.
In this research question, we focus on \nIndirectComp breaking dependency updates classified as \indirectCategory.

\begin{table}[htp]
\caption{Changes in indirect dependencies that trigger breaking updates}
\centering
\rowcolors{2}{gray!10}{white}
\begin{tabular}{cc}
\toprule
Indirect dependency modification  &  \makecell{Number of breaking updates} \\ 
\midrule
  REMOVED DEP. & 25\\
  UPDATED DEP.  &  7\\
  ADDED DEP.  &  6 \\
  
\bottomrule
 
\end{tabular}
\label{tab:indirect_modification}
\end{table}

\autoref{tab:indirect_modification} presents the three types of changes that we observe a number of breaking updates because indirect dependencies are removed, modified, or added in the dependency update.
A total of 27 breaking updates fail due to indirect dependencies being removed in the new version of the updated direct dependency.
The addition of indirect dependencies caused 6 breaking dependency updates, and 7 breaking updates out of \nIndirectComp fail due to the update of an indirect dependency.
All those dependency updates introduce changes in the indirect dependencies.
 
For 38 breaking updates, more than one indirect dependency is updated.
In 37/38 breaking updates, only one of the updated indirect dependencies causes the failure.
One breaking update is due to breaking changes in two indirect dependencies.

Adding a new indirect dependency can result in a breakage, when the added dependency creates a version conflict with the same dependency that is present elsewhere in the dependency tree, but with a different version. In this case, the Maven dependency resolution algorithm selects the version from the update. 
For example, by updating the \texttt{asto-core} dependency from \texttt{v1.13.0} to \texttt{v1.15.3} in the \texttt{docker-adapter} project, the new transitive dependency \texttt{hamcrest-core} is added. Yet, \texttt{hamcrest} is declared as a direct dependency of \texttt{docker-adapter}.
\texttt{hamcrest-core} is a higher version of \texttt{hamcrest}, which contains the same class structure as \texttt{hamcrest}.
The Maven algorithm resolves to use \texttt{hamcrest-core} instead of \texttt{hamcrest}, modifying the project's classpath, causing the breakage.
For example, the \texttt{StringContainer} class is contained in both dependencies.
Yet, the class constructor is not the same in both dependencies, resulting in incompatibilities that cause \textit{constructor cannot be applied to given types} error during compilation build.

Removing indirect dependencies in the new version of an updated dependency can considerably impact the client application. 
Recall that in Java, it is perfectly normal for application developers to invoke functions from indirect dependencies. While this is not a recommended practice with respect to encapsulation, we have seen this happen. In this case, the removal of the indirect dependency breaks the application's build because it relies of non-existent code. For example, when the developer of \texttt{code-coverage-api-plugin} project updates the \texttt{acceptance-test-harness} dependency from version \texttt{5631.v2dcb\_f66e58f7} to version \texttt{5588.vd13b\_52985008}, a total of 27 indirect dependencies are removed in the new version.
The removal of the \texttt{htmlunit} dependency causes the compilation of the client code to fail because it is not possible to access constructs that belong to the removed dependency, causing \textit{cannot find symbol} errors. For future work on mitigating this problem, we a static analysis for identifying client code using indirect dependencies.

We consider that an indirect dependency is updated if its version changes and it involves changes to the API or behavior.
If a modified construct from the indirect dependency is used in the client application, this can lead to compilation errors that result in a breaking update.
For the client, the use of indirect dependency constructs that are modified creates a breakage.
For example, updating the \texttt{maven-surefire-common} dependency from version \texttt{3.0.0-M5} to version \texttt{3.0.0-M7} in the \texttt{flacoco} project fails to compile the client application code.
A failure occurred because the \texttt{getWildCard} method can not be found.
This method belongs to the indirect dependency \texttt{surefire-api}, which was updated under the hood from version \texttt{3.0.0-M5} to version \texttt{3.0.0-M7}.
The method was removed in the new version of the \texttt{surefire-api} dependency which causes \textit{cannot find symbol} errors.

\begin{tcolorbox}[boxrule=1pt,arc=.3em, left=4pt, right=4pt]
\textbf{Answer to \ref{rq: transitive}}:
In the field, indirect dependencies that are changed as a ripple of updating one direct dependency can trigger a breaking update.
We observe that additions, deletions, and modifications of indirect dependencies create various scenarios where the change propagates as a compilation error in the client application. There is a need for research to study this problem which happens a lot in the field. We believe that dedicated tooling would help developers 1) to ensure that no function from indirect dependencies are called and 2) to help developers understand and fix the indirect breakage when this happens. 
\end{tcolorbox}

\subsection{RQ3 (User Study)}
\label{quality}

In this research question, we evaluate the quality of automatically generated explanations.
To evaluate the quality of the explanations, we survey a group of researchers in computer science.
The survey allows us to collect feedback about the generated explanations in terms of quality, usefulness, and accuracy.
We evaluate 10 generated explanations in total covering all categories, allowing for inclusive and diverse feedback.
To illustrate the result, we use the explanation shown in \autoref{fig:explanation}  generated by \tool and evaluated by the participants.

Identifying the causes of breakages is a challenging task for developers.
Our explanations include a specific section to address that problem, pointing to the cause of the breaking update. 
The explanation in \autoref{fig:explanation}  shows that the cause of the breaking update is the modification of the \texttt{between} method.
For 8 cases in our survey, 4 of the participants felt that identifying the cause of a breaking update was \say{very difficult} or \say{extremely difficult} when only having access to the Maven logs.
However, when we provided the generated explanation only 1 participant reported difficulty in identifying the cause of the breakage.
Additionally, 4 participants reported that detailed explanations made it much easier to identify the root cause of the problem, providing a valuable foundation for addressing its resolution.
\begingroup
\setlength{\tabcolsep}{12pt}
\renewcommand*{\arraystretch}{1.8}
\addstackgap[4pt]{
\begin{tabular}{!{\color{black}\vrule width 2 pt}p{7.5cm}}
\emph{It explained exactly where the bugs occurred in the program and provided a more visualized explanation of what had happened.}\\
\end{tabular}
}
\endgroup
\tool's explanations significantly improve the understanding of the cause of breaking updates.

Regarding the usefulness of the explanations, all participants rated them as \say{very useful} for resolving issues in the compilation process.
One participant highlighted the value of the explanations, claiming the explanations would significantly reduce the time needed to analyze breaking updates.
The qualitative feedback revealed several strengths of our novel approach.
The survey participants appreciated the clarity and precision of the explanations, as well as their ability to contextualize the problem.
The participants shared that the suggestions helped to resolve the failure. 

The survey shows that the explanations are accurate, which can help reduce the time required to resolve the error.
One developer mentions that adding icons to the sections covering the constructs causing the errors helps to focus on the important sections of the explanation.
On the other hand, the participants agree that grouping all the information related to the breaking update together helps to understand the error better.
This positive feedback validates some key design decisions of \tool.

\begingroup
\setlength{\tabcolsep}{12pt}
\renewcommand*{\arraystretch}{1.8}
\addstackgap[4pt]{
\begin{tabular}{!{\color{black}\vrule width 2 pt}p{7.5cm}}
\emph{The idea of providing a detailed explanation of the error shown in the Maven compilation process is already very amazing.}\\
\end{tabular}
}
\endgroup

However, some weaknesses were also identified.
Some respondents mentioned that, in certain cases, explanations could benefit from further customization of the information presented.
For example, the identified constructs could be represented with a simple name instead of the fully qualified name, improving the construct's readability and identification in the client's code.
\begin{figure}[hpt]
    \centering
    \includegraphics[width=.90\linewidth]{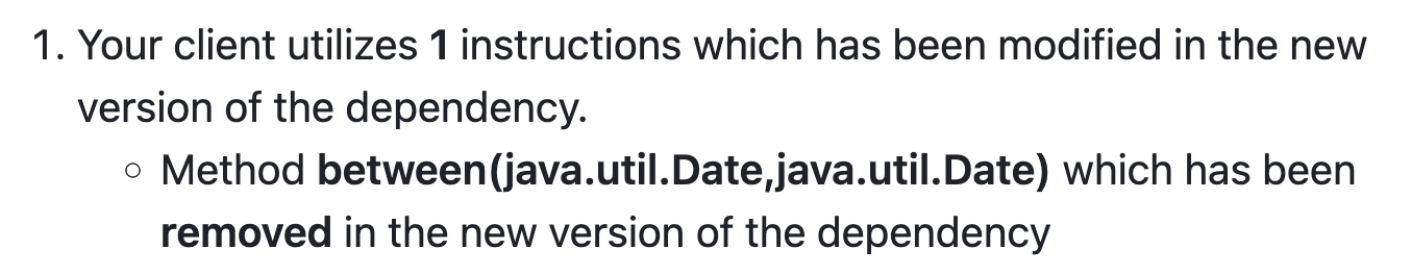}
    \caption{Example of construct that causes breaking update.}
    \label{fig:method}
\end{figure}
In the example of \autoref{fig:method}, participants say that the names of the parameters that are passed to the method can be represented in a more human-readable text.
Another weakness identified was the inclusion of error lists extracted from Maven logs.
One developer pointed out that having duplicate information seemed unnecessary.

\begin{tcolorbox}[boxrule=1pt,arc=.3em, left=4pt, right=4pt]
\textbf{Answer to \ref{rq: explanations}}:
In the user study, the surveyed developers appreciated the clarity of explanations generated by \tool. They noted the usefulness of concrete feedback given to fix the breaking update.
Overall, the use study gives evidence that \tool generates actionable explanations, which are effective at helping developers finding the root cause of a breaking update. 
\end{tcolorbox}

\section{Threats to Validity}

The first threat relates to the possible imprecision of log parsing to identify the lines of code that are responsible for compilation errors.
To mitigate this threat, we performed an exhaustive comparison between the results obtained using \tool and a manual analysis of the build logs to validate the accuracy of the approach, fully validating the implementation.

A similar internal threat concerns the detailed categorization of compilation errors through pattern matching in the logs.
To address this threat, we performed a manual verification of the classifications used by \tool to generate the explanations, comparing the results with manual analysis.
However, the risk remains that we miss special patterns for some less frequent types of errors.

To minimize threats to external validity, we applied \tool in 82 different projects containing real breaking dependency updates, ensuring the generalization of the results.
In addition, our user study confirmed the usefulness of the explanations to identify the causes of the broken update beyond automated evaluation, mitigating the applicability threat.

\section{Related Work}
\label{sec: related_work}

In the following section, we discuss the state of the art in two areas that address the problem of breaking dependency updates: build breakage assistance and breaking changes.

\subsection{Build Breakage Assistance}
\label{subs: build_breakage}

In software development, build breakages occur due to various factors, including changes in dependencies. Addressing these build breakages is crucial for maintaining the stability and functionality of the software.

Vassallo \etal \cite{VassalloEveryResolution} propose BART to summarize Maven build failures and suggest solutions from StackOverflow, aiming to assist developers in understanding and fixing build breakages.
Zhang \etal introduce BuildSonic \cite{Zhang2022BuildSonic:Builds} to address configuration problems in continuous integration (CI) processes. The study presents a comprehensive catalog of CI performance configuration issues, detailing each element including its description, solution, and identification method.
Al-Kofahi \etal \cite{AlKofahi2014FaultMakefiles} detect bugs that trigger build issues in Makefile.
Silva \etal \cite{DaSilva2022BuildWild} investigate the frequency, structure, and resolution patterns of build conflicts arising from collaborative software development, based on analysis of 451 open-source Java projects' merge scenarios and Travis build logs.
HireBuild \cite{Hassan2018HireBuild:Scripts} proposes automatic patch generation for Gradle build scripts, using fix patterns derived from historical fixes and making recommendations based on build log similarity. HireBuild successfully fixed 11/24 (45\%) build failures in a time comparable to manual fixes.
Lou \etal \cite{Lou2020UnderstandingPatterns} investigate the resolution of build failures in Maven, Ant, and Gradle, analyzing 1,080 Stack Overflow issues and identifying effective fix patterns. It highlights the prevalence of issues in build scripts related to plugins and dependencies, underscoring the need for nuanced approaches to mitigate the challenges developers face in resolving build failures.
Li \etal \cite{Li2022Nufix:Maze} study the challenges of the Dependency Maze (DM) faced by developers in the .NET ecosystem by proposing NuFix, an automated technique that utilizes empirical studies of real DM issues to generate high-quality fixes adopted by popular .NET projects.

\tool differs from the previous approaches, as it analyzes the logs generated during the build process and then analyzes the source code of the application to locate the lines that cause the failure.
This novel combination of code analysis and log analysis allows us to identify the client elements affected by the new dependency version.

\subsection{Studies of Breaking Changes}
\label{subs: studies_bc}
Research on the problem of breaking changes in software dependencies and its implications for development have been addressed through different studies.

Yamaoka \etal \cite{YamaokaComparingIncompatibilities} propose a technique using Merkle trees to identify incompatibilities in client-side library updates.
This method compares execution traces before and after the update to identify possible library methods that could cause incompatibilities. 
PyDFix \cite{Mukherjee2021} detects dependency errors in Python compilations. PyDFix is evaluated on two bug datasets, BugSwarm and BugsInPy, identifying a high proportion of builds as unreproducible due to dependency errors.
Bogart \etal \cite{Bogart2021WhenEcosystems} explore the impact of disruptive updates on open source projects, investigating methods and processes used by software development communities to manage and adapt to major changes, analyzing platforms (Eclipse, NPM, and CRAN).
Supatsara \etal\cite{Wattanakriengkrai2023LessonsEcosystems} investigate the acceptance rates of insecure dependency updates across 1,500 JavaScript npm libraries categorized into Top-500, Middle-500, and Bottom-500 tiers, revealing higher acceptance rates in lower-tier libraries.
CompCheck \cite{Zhu2023Client-SpecificDiscovery} analyzes how a dependency update affects client behavior. CompCheck generates a knowledge base from compatibility issues between clients and their updates. A total of 202 sites in 37 open-source projects are examined by CompCheck, identifying incompatibilities in 76 of them.
Jayasuriya \etal \cite{Jayasuriya2023UnderstandingWild} study the impact of breaking changes in client projects, analyzing 18,415 Maven artifacts from Java projects. The study reports that 11.58\% of dependency updates introduce breaking changes affecting clients, caused by changes in direct and indirect dependencies.
In our study, we identified that 38 out of 243 updates of break dependencies were due to changes in indirect dependencies. 
This corroborates the findings of Jayasuriya highlighting the impact of indirect dependencies in breaking dependency updates.

The closest work to our approach is Maracas\cite{Ochoa2022BreakingStudy}. Maracas analyses the changes between two versions of a Java library using a static analysis approach based on Java bytecode. 
Maracas determines if the changes introduced by the new version of the dependency are used in the client's code, providing a detailed understanding of the implications of library updates on client code. 
\tool focuses on identifying the changes introduced in the new version of the dependency and their relationship with the compilation failures detected in the build process of the client application, while Maracas identifies these changes through static analysis of client code. 
Our approach accurately identifies which lines are affected by the changes introduced, unlike Maracas, which identifies lines that are not invoked in the build process. 
While Maracas only analyzes changes between two versions of a dependency and how they affect client code, \tool also analyzes how changes in indirect dependencies defined in the new version of the dependency influence client behavior.
\tool analyzes if the elements causing the breaking dependency update belong to an indirect dependency that has been modified, removed, or added in the new version of the dependency being updated.

\section{Conclusion}

In this paper, we have introduced \tool, a novel technique that combines log and dependency analysis to automatically generate explanations for breaking dependency updates.
\tool provides a detailed categorization of the causes of compilation errors triggered by dependency updates.
We are the first to report quantitative and qualitative findings on the main causes of compilation error due to breaking updates: API changes in direct and indirect dependencies, incorrect build configurations and incompatibilities between Java versions required by the client and dependencies.
To our knowledge, \tool is the first technique that highlights and mitigates the common problem in practice of indirect dependencies causing breaking updates.
Overall, \tool is an essential missing tool in the developer toolbox, facilitating the understanding and resolution of breaking updates and reducing the time and effort spent in dependency-related software engineering problems.

As future work, we will investigate the explanation of breaking dependency updates occurring when more than one dependency is updated at the same time, in a single pull request, a scenario already supported by bots like Renovate.

\section*{Acknowledgment}
\label{sec:ack}

This work was supported by the CHAINS project funded Swedish Foundation for Strategic Research (SSF), the WebInspector project funded by the Swedish Research Council (VR), as well as by the Wallenberg Autonomous Systems and Software Program (WASP) funded by the Knut and Alice Wallenberg Foundation.

\bibliographystyle{IEEEtran}
\IEEEtriggeratref{17}
\bibliography{reference}

\end{document}